\documentclass[useAMS,usenatbib,usegraphicx]{mn2e} 
\usepackage{amssymb} 
\usepackage{times} 
\usepackage{aas_macros} %needed to define journal macros generated by 
\usepackage[usenames]{color} 
\usepackage{amsmath}
\usepackage{color}
\usepackage{ulem}
\usepackage{hyperref}

\usepackage{lineno}
\title[First Tri-Axial Pulsator]{TIC 184743498: The First Tri-Axial Stellar Pulsator}

\vspace{0.1cm}

\author[Zhang et al.]{Valencia Zhang$^1$\thanks{E-mail: vzhang25@andover.edu}, Saul Rappaport$^2$, Rahul Jayaraman$^2$, Donald W. Kurtz$^{3,4}$,  \newauthor Gerald Handler$^{5}$, James Fuller$^{6}$, and Tamas Borkovits$^{7,8,9,10,11}$
\\
$^1$Phillips Academy, Andover, MA, 01810, USA \\
$^{2}$Department of Physics, and Kavli Institute for Astrophysics and Space Research, M.I.T., Cambridge, MA 02139, USA\\
$^{3}$Centre for Space Research, North-West University, Mahikeng 2745, South Africa\\
$^{4}$Jeremiah Horrocks Institute, University of Central Lancashire, Preston PR1 2HE, UK\\
$^{5}$Nicolaus Copernicus Astronomical Center, Polish Academy of Sciences, ul. Bartycka 18, 00-716, Warszawa, Poland\\
$^{6}$TAPIR, Mailcode 350-17, California Institute of Technology, Pasadena, CA 91125, USA\\
$^{7}$Baja Astronomical Observatory of University of Szeged, H-6500 Baja, Szegedi \'ut, Kt. 766, Hungary\\
$^{8}$ HUN-REN--SZTE Stellar Astrophysics Research Group, H-6500 Baja, Szegedi \'ut, Kt. 766, Hungary\\
$^{9}$Konkoly Observatory, HU-REN Research Centre for Astronomy and Earth Sciences,  H-1121 Budapest, Konkoly Thege Mikl\'os \'ut 15-17, Hungary\\
$^{10}$ELTE E{\"o}tv{\"o}s Lor\'and University, Gothard Astrophysical Observatory, Szent Imre h. u. 112, 9700 Szombathely, Hungary \\
$^{11}$HUN-REN-ELTE Exoplanet Research Group, H-9700 Szombathely, Szent Imre h. u. 112, Hungary \\
}
\date{Accepted XXX. Received YYY; in original form ZZZ}

\pubyear{2020}

\begin{document}
\label{firstpage}
%\pagerange{\pageref{firstpage}--\pageref{lastpage}}
\maketitle

\begin{abstract}
We have discovered a $\delta$ Scuti pulsator in a tight binary (P = 1.053\,d) with nine pulsation modes whose frequencies are between 38 and 56\,d$^{-1}$. Each of these modes exhibits amplitude modulations and $\pi$-rad phase shifts twice per orbital cycle.  Five of these modes exhibit amplitude and phase shifts that are readily explained by dipole pulsations along an axis that is aligned with the binary's tidal axis.  The novelty of the system lies in the remaining four pulsation modes, which we show are dipole pulsations along an axis that is perpendicular to both the tidal axis and the binary's orbital angular momentum axis.  There are additionally two pulsation modes whose amplitudes and phases do not change significantly with orbital phase; they are explained as dipole modes along an axis aligned with the orbital/rotation axis.  Hence, we propose that TIC 184743498 is a tri-axial pulsator, the first of its kind.

%Asteroseismology provides a window into the interior of pulsating stars. Its application requires knowledge of the types of pulsations observed. Modulation of the pulsation amplitudes and phases of oscillating stars in oblique pulsators, where the pulsation mode is viewed from varying aspect with rotation or orbital motion in a binary, provides this information. Here, for the first time, we report the discovery of a tidally distorted binary star pulsating around three different, orthogonal pulsation axes, providing rich asteroseismic information. Five of its pulsation modes are dipole pulsations along the binary’s tidal axis, four modes are dipole pulsations along an axis perpendicular to both the tidal and the orbital angular momentum axis, and at least one mode is a dipole aligned with the orbital/rotation axis. We show theoretically how such alignments occur. Our novel observational and theoretical results expand the application of asteroseismic techniques to several classes of pulsators in close binary systems.
\end{abstract} 


\section{Introduction}
\label{sec:intro}

Studies of the photometric variability of stars due to stellar pulsations have been carried out for over a century. For instance, the variability of the bright early F-type star, Delta Scuti ($\delta$ Sct), which lends its name to an entire class of pulsating stars, was discovered in radial velocities by \citet{1900ApJ....12..254C}. They found a range of about 10\,km~s$^{-1}$ in 7 measurements made from 1899 to 1900. \citet{1935PASP...47..232F} and \citet{1935PASP...47..231C} noted sinusoidal photometric variability in $\delta$~Sct stars with maximum light at minimum radial velocity over a period of 0.193\,d, and concluded that the star is not a spectroscopic binary. \citet{1973A&A....23..221B}, \citet{1979PASP...91....5B,breger00}, and \citet{2022ARA&A..60...31K} summarize more recent work on this class of star.

$\delta$~Sct stars, along with almost all other classes of pulsating variables, are observed to have nonradial modes. Unlike the symmetrical radial pulsations of the dominant modes of Cepheid variables (as described in, e.g., \citealt{1926ics..book.....E}), nonradial modes have a pulsation axis. Originally, it was implicitly assumed that a star's pulsation axis coincides with its rotation axis, as it is about this axis that most stars deviate from spherical symmetry. It is still the case in modern asteroseismology \citep{2010aste.book.....A} that the starting point for modeling and inference is the assumption of the alignment of the pulsation and rotation axes.

However, stars can be distorted from spherical symmetry by other effects, primarily the presence of strong, global magnetic fields, and by tides arising from close companions. Therefore they may have pulsation axes not aligned with the rotational axis. The first stars found to pulsate about an axis other than the rotation axis were the rapidly oscillating Ap (roAp) stars \citep{1982MNRAS.200..807K}, which have primarily dipole magnetic fields that are inclined to the rotation axis. The combined effects of the rotation and magnetic fields lead to a pulsation axis that is inclined to the rotation axis and possibly offset from the stellar center. As the star rotates, the pulsation mode is then seen from varying `latitudinal view angles'\footnote{Consider a star with a spin axis defined by $\hat{s}$ and a pulsation axis, $\hat{p}$ lying at an obliquity angle $\beta$ with respect to the spin axis, and is viewed by a distant observer along a direction $\hat{v}$, such that $\hat{v} \cdot \hat{s} \equiv \cos i$, and $i$ is the inclination angle.  We define an angle $\Theta$, the latitudinal view of the pulsation axis, as $-\hat{p} \cdot \hat{v}$.  If $\hat{p}$ revolves about $\hat{s}$ by an angle $\phi$ due to either the rotation of the star or the orbital motion in the case of a tidally tilted pulsation axis, then it is straightforward to show that: $\cos \Theta = \cos i \cos \beta + \sin i \sin \beta \cos \phi$. If $\beta =0$, i.e., no obliquity, then $\Theta = i$, a constant.}, $\Theta$, leading to modulation of both the observed pulsation amplitude and phase. This phenomenon is known as oblique pulsation. 

\citet{1982MNRAS.200..807K} introduced the oblique pulsator model for the roAp stars, which has been refined over the last 40 years for asteroseismic inference.  \citet{2021MNRAS.506.1073H} provides the most up-to-date analysis of a large sample of roAp stars that were observed during the first two years of the Transiting Exoplanet Survey Satellite's ({\it TESS}) mission \citep{2015JATIS...1a4003R}.  A natural extension of the oblique pulsator model is to stars in close binary systems, wherein the pulsation axis could align with the tidal axis (the line joining the two stars). Since $\delta$~Sct stars are common upper main-sequence pulsators that are often found in close binary systems, it was no surprise that such ``tidally tilted pulsators'' (TTPs) were found in {\it TESS} data.
  
The first such system that exhibited TTPs, HD~74423, is a pair of nearly identical chemically peculiar $\lambda$\,Boo stars that nearly fill their Roche lobes in a 1.58-d binary \citep{2020NatAs.tmp...45H}. Only one of the stars is pulsating and in only one mode, which is unusual for a $\delta$~Sct star. The pulsation mode in HD~74423 is a highly distorted dipole mode that is, remarkably, largely confined to one hemisphere of the star. This discovery provided the impetus for the theoretical groundwork for understanding the interaction of pulsation and tidal distortion in close binary stars  \citep{2020MNRAS.tmp.2716F}. Such pulsations can provide significantly more information about the mode geometry than can be gleaned from `normal' nonradial modes, and allows for the identification of the mode degree, $\ell$, and azimuthal order, $m$ (see, e.g., \citealt{reed05}). 
 
The second TTP found was CO~Cam, a marginal Am star in a 1.27-d binary with an undetected companion. It pulsates in at least four tidally tilted modes  \citep{2020MNRAS.494.5118K}. Then the third discovery was TIC~63328020, which showed, for the first time, a tidally tilted sectoral ($|m| = \ell$) dipole mode \citep{rappaport21}. Previously, all identified modes in roAp stars had been zonal ($m=0$) distorted dipole and quadrupole modes; this discovery allowed for further study of non-axisymmetric tidally tilted pulsations.  The fourth discovery was HD\,265435, a subdwarf B star which has been stripped of its H-rich envelope; this star has over 30 modes, of which at least 25 are tidally tilted \citep{jayaraman22}. 

There has also been recent interest in less extreme examples of this phenomenon (the ``tidally perturbed pulsators,'' see, e.g., \citealt{2020MNRAS.497L..19S,2023A&A...671A.121V,2023A&A...670A.167J}).  Because our understanding of mode excitation, mode selection, and mode trapping in all pulsating stars remains rudimentary, discoveries of tidally tilted (and perturbed) pulsators with more, and different, kinds of pulsation modes provides further constraints on developing theory and asteroseismic inference. 

However, the oblique pulsator model has been unable to explain certain observations of roAp stars. For instance, \citet{2011MNRAS.414.2550K} concluded that the two pulsation modes of the roAp star KIC~10192926 had different pulsation axes, both of which were inclined to the rotation axis. For another roAp star HD~6532, \citet{2020ASSP...57..313K} found that the single pulsation mode appeared to have a different pulsation axis when observed through the red {\it TESS} filter, compared to the pulsation axis derived from ground-based $B$-band observations. Because these two filters sample different atmospheric depths, perhaps the mode in this star has a complex three-dimensional geometry; this result may imply that different pulsation modes in a single star may be trapped in mode cavities with different pulsation axes. This effect remains unstudied.

In this work, we report the discovery of TIC~184743498, a close binary system whose primary star exhibits $\delta$~Sct pulsations over the frequency range $40-55$\,cycles d$^{-1}$. This star exhibits nonradial modes with pulsation axes aligned with the rotation axis, the tidal axis, and a third axis perpendicular to both of those. Thus, it is a tri-axial pulsator, the first of its kind. First, Section \ref{sec:discovery} describes how this source was discovered. Then, Section \ref{sec:data} discusses the data we used in addition to the photometry from {\it TESS}.  In Section \ref{sec:stellarparam}, we use the available observational data on this system to make robust estimates of the stellar parameters of the two binary stars, as well as a tertiary star, in TIC 184743498.  The pulsation spectrum is discussed in Section \ref{sec:pulsations}, and the amplitudes and phases of these pulsations are reconstructed as a function of orbital phase in Section \ref{sec:reconstruct}.  The expected density of radial pulsation modes for TIC 184743498 is estimated in Section \ref{sec:radial_modes}. We present in Section \ref{sec:models} a rudimentary perturbation model that can explain in a natural way how this star could have three different orthogonal pulsation axes.  We summarize our results in Section \ref{sec:sumconclude}, and look at future applications of our findings.

\section{Discovery}
\label{sec:discovery}

The first two TTPs discovered (\citealt{2020NatAs.tmp...45H}; \citealt{2020MNRAS.494.5118K}) were found serendipitously by visual examination of millions of {\it TESS} light curves from full frame images \citep{2022PASP..134g4401K}.  Since then, we have focused on {\it TESS} targeted stars with 120-s cadence calibrated light curves generated by the Science Processing Operations Center (SPOC; \citealt{spoc_proceedings}). One of us (RJ) carried out a preliminary search for periodicities in all of the approximately 20,000 SPOC sources processed for each sector.  A single summary sheet is generated for the approximately 1,000 stars from each sector that exhibit significant periodicity.  The summary sheet includes plots of the raw light curve, the Fourier transform (FT) with various scalings, the inferred principal period, and a light curve folded about this period. The summary sheets also provide properties for each source taken from the {\it TESS} Input Catalog (TIC v8.0; \citealt{2019AJ....158..138S}), including the effective temperature, sky location, and estimated radius. Some of us then visually examine these summary sheets to search for interesting periodic stellar phenomena, including stellar pulsations in binary systems.  

\begin{figure*}
\centering
\includegraphics[width=0.9\linewidth,angle=0]{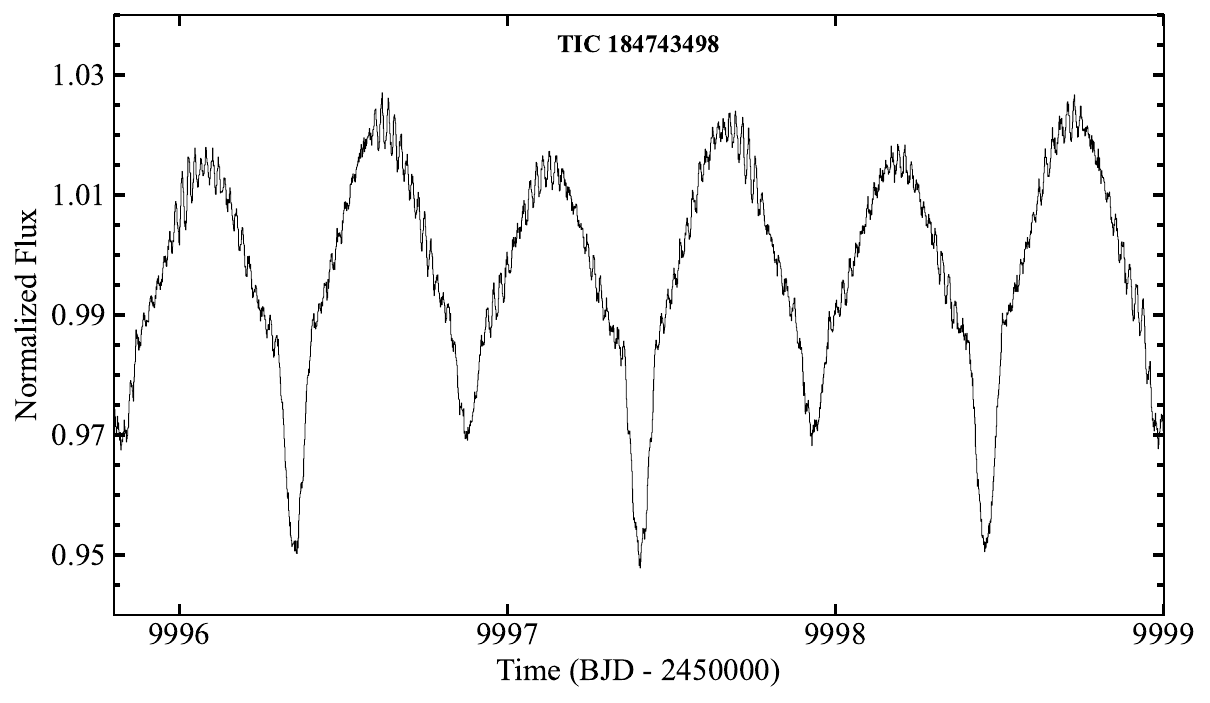} 
\caption{A portion of the Sector 62 TIC~184743498 light curve showing the eclipses, ellipsoidal light variations, and prominent pulsations.  The {\it TESS} cadence for this sector was 120\,s.}
\label{fig:lc1}
\end{figure*}

As a part of a continuing effort to identify TTPs, VZ visually inspected summary sheets for sources from the recent {\it TESS} Sectors 56 -- 62, specifically looking for stars in binary systems whose FT displayed peaks at higher frequencies (typically $\gtrsim$\,15\,d$^{-1}$).  During visual inspection of the summary sheets from Sector 62, she found an interesting set of pulsations in TIC~184743498. In particular, she noticed high frequency pulsations between $40-55$\,d$^{-1}$ in a $\sim$ 1\,d eclipsing close binary.   Fig.~\ref{fig:lc1} shows a small portion of the {\it TESS} light curve for TIC~184743498 covering about 3\,d during Sector 62.  Shallow eclipses, ellipsoidal light variations (ELV), and prominent pulsations are salient features of the raw light curve. 

\begin{figure}
\centering
\includegraphics[width=1.00\columnwidth,angle=0]{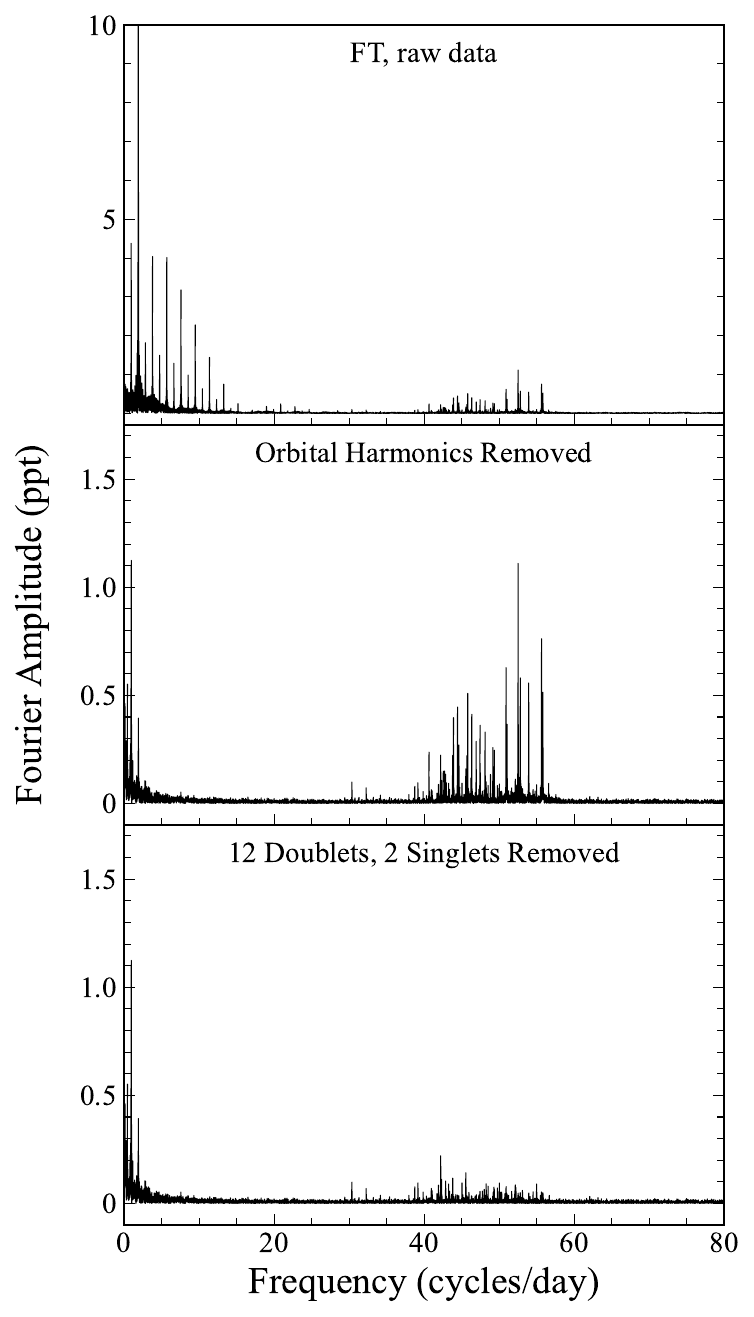}	
\caption{The Fourier amplitude spectrum of TIC 184743498. The panels highlight the amplitude spectrum after various stages of cleaning. The top panel shows the spectrum of the raw {\it TESS} light curve; the middle panel represents the spectrum following the removal of 30 orbital harmonics, while the bottom results from a further cleaning of the 13 pulsation modes discussed in the paper and listed in Table \ref{tbl:frequencies}. Note the changes in amplitude scale between the upper and lower panels.}
\label{fig:ft1}
\end{figure}

The top panel of Fig.~\ref{fig:ft1} shows the Fourier amplitude spectrum of data from {\it TESS} Sectors 61 and 62 before any filtering.  The FT clearly displays numerous orbital harmonics and a rich set of pulsations from TIC~184743498.  The lower panels show the FT after the orbital harmonics have been removed, and then after the largest amplitude pulsations have been cleaned out.  This `cleaning' process for the Fourier transform is explained later in Section \ref{sec:pulsations}. 

  % Figure 1.  Snippet of the lightcurve.

  % Figure 2. FT panels.

\section{Data Availability}
\label{sec:data}

\subsection{TESS Data}
\label{subsec:tessdata}

TIC~184743498 was observed by {\it TESS} at 30-min cadence in Sector 8 (from 2019 May 16 to June 13), at 10-min cadence in Sectors 34 and 35 (2021 January 14 to March 6), 200-s cadence in Sector 61 (2023 January 18 to February 12), and 2-min cadence in Sector 62 (2023 February 12 to March 10).  This target was selected for 2-min cadence observations because it was part of the {\it TESS} Candidate Target List (CTL v8.01), a collection of bright ($T < 13$) stars in the {\it TESS} field of view that could host exoplanets. We accessed the data using the Python package \texttt{lightkurve} \citep{lightkurve}. We used the Sector 61 and 62 data (200-s and 2-min cadence) for our principal pulsation analysis, but also utilised the Sector 34 and 35 data to confirm our findings based on the later sectors, and to better determine the orbital period. The Sector 61 and 62 data, when combined, span 51\,d and contains 28\,700 data points. 

\subsection{Other Data}
\label{subsec:otherdata}

We also made use of archival data from Gaia Data Release 3 \citep{gaia_dr3_full}, the Mikulski Archive for Space Telescopes (MAST)\footnote{\url{https://mast.stsci.edu/portal/Mashup/Clients/Mast/Portal.html}}, the All-Sky Automated Survey for SuperNovae (ASAS-SN; \citealt{shappee14}; \citealt{kochanek17}), and the online VizieR SED viewer (\citealt{ochsenbein00})\footnote{\url{http://vizier.cds.unistra.fr/vizier/sed/}}.  The basic archival parameters of TIC~184743498 from Gaia and MAST are summarized in Table \ref{tbl:mags}.  We used the ASAS-SN data primarily to obtain an accurate and independent value of the orbital period over a baseline of $\sim$10\,yr.  The spectral energy distribution (SED) data were obtained from VizieR; these points were used in conjunction with other data to estimate the properties of the two stars in the binary in Sect~\ref{sec:stellarparam}.

Fortuitously, TIC~184743498 is one of the relatively rare double-line spectroscopic binaries in the Gaia data set \citep{gaia_rvs, gaia_hot_rvs}. The Gaia $K_1$ and $K_2$  velocities, also listed in Table \ref{tbl:mags}, are 119.3 and 154.9 km~s$^{-1}$, respectively.

\subsection{Orbital period determination}
\label{sec:period}

We determined the orbital period of TIC~184743498 using several different approaches. 

First, using all five sectors of {\it TESS} data, we performed a Box Least Squares (BLS) analysis\footnote{\url{https://exoplanetarchive.ipac.caltech.edu/cgi-bin/Pgram/nph-pgram}} \citep{kovacs02}.  This method resulted in a period of 1.053\,234\,d.  We also performed a BLS transform of the nearly 7,000 ASAS-SN archival flux points spanning 11\,yr.  The period derived from that data set is 1.053\,238\,d.  The Gaia archives provide a period of $1.053\,245 \pm 0.000\,014$\,d, based on their RV solution to the orbit.  These results are summarized in Table \ref{tbl:orbit}.

\begin{figure}
\centering
\includegraphics[width=1.00\columnwidth,angle=0]{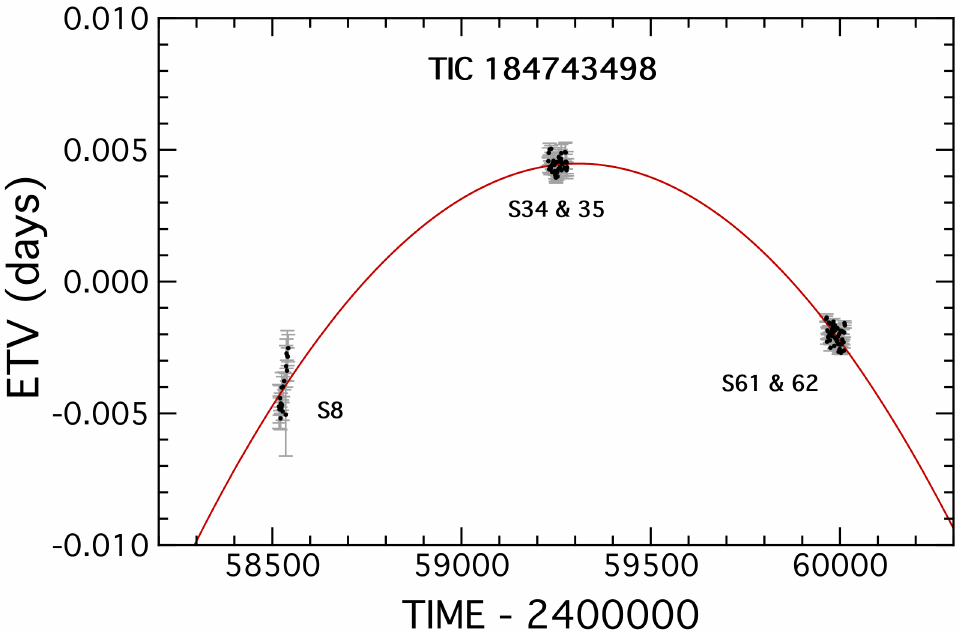} 	
\caption{ETV curve based on 107 primary eclipse times from {\it TESS} Sectors 8, 34, 35, 61, and 62. The reference period for this analysis was 1.053\,240\,0 d.  The non-linear behavior is significant at the 80 $\sigma$ level.}
\label{fig:etv}
\end{figure}  % Figure 3. ETV curve.

Additionally, we generated an eclipse timing variations (ETV) curve from the {\it TESS} data.  The results are shown in Fig.~\ref{fig:etv}.  The reference period used to compute this ETV curve was 1.053\,240 d. The residuals from a linear ephemeris are nicely fitted with a quadratic term which can be described as a fractional rate of change in the orbital period of $\dot P_{\rm orb}/P_{\rm orb} \simeq -(1.030 \pm 0.013) \times 10^{-5}$ yr$^{-1}$.  This curvature in the ETV curve could represent either orbital decay or a portion of a long-period orbit about a third body.  Given that Gaia additionally reports an `acceleration' solution for this object, this implies a non-linear component to the proper motion \citep{gaia_acc}. This is consistent with orbital motion about a third body rather than orbital decay, and we adopt the former hypothesis.  

	The mass of the third body necessary to induce this curvature in the ETV curve can be estimated to be in the range of $\sim$0.7-1.3 M$_\odot$ if the outer orbit has a period in the range $\sim$2000-6000 d.  Such an orbital period would be consistent with (i) the fact that Gaia detects an accelerated astrometric motion, (ii) the duration of the Gaia observations, and (iii) the parabolic shape of the ETV curve, as opposed to a sinusoidal shape.  With this possible mass range for the third star, we have to allow for the fact that it may make a non-negligible contribution to the system light. 

\begin{table}
\centering
\caption{Properties of the TIC~184743498 System}
\begin{tabular}{lc}
\hline
\hline
Parameter & Value   \\
\hline
RA (J2000) (h m s)& 08:31:11.649  \\   %below all from MAST
Dec (J2000) ($^\circ \ ^\prime \ ^{\prime\prime}$) &  -39:06:45.41 \\ 
$T$$^a$ & $9.356 \pm 0.006$ \\  %tess magnitude 
$G$$^b$ & $9.546 \pm 0.001$  \\ %gaia G magntiude
$G_{\rm BP}$$^b$ & $9.6905 \pm 0.0035$  \\ %gaiabp
$G_{\rm RP}$$^b$ & $9.3181 \pm 0.0038$  \\ %gaiarp
$B^a$ & $	12.57 \pm 0.10$ \\ %bmag
$V^a$ & $9.913 \pm 0.012$ \\%vmag
$J^a$ & $8.961 \pm 0.021$   \\ %jmag
$H^a$ & $8.883 \pm 0.026$  \\ %hmag
$K^a$ & $8.78 \pm 0.021$ \\ %kmag
W1$^c$ & $8.767 \pm 0.023$ \\ %w1
W2$^c$ & $8.782 \pm 0.021$  \\ %w2
W3$^c$ & $8.753 \pm 0.025$  \\ %w3
W4$^c$ & $8.4 \pm 0.3$  \\ %w4
$K_1$ (km~s$^{-1}$)$^b$ & $154.9 \pm 9.5$ \\ %k1, k2, gamma from vizier
$K_2$ (km~s$^{-1}$)$^b$ & $119.3 \pm 2.9$ \\
$\gamma$ (km~s$^{-1}$)$^b$ & $17.2 \pm 2.0$\\
Distance (pc)$^b$ & $ 393.3 \pm 	4.8$  \\   %from mast
$\mu_\alpha$ (mas ~${\rm yr}^{-1}$)$^b$ & $-5.11 \pm 0.03$   \\  
$\mu_\delta$ (mas ~${\rm yr}^{-1}$)$^b$ &  $+8.91 \pm 0.04$   \\ 
\hline
\label{tbl:mags}  % Table 1 of star properties.
\end{tabular}

{\bf Notes.}  (a) MAST (\url{https://mast.stsci.edu/portal/Mashup/Clients/Mast/Portal.html}).  (b) Gaia DR3: \citep{gaia_dr3_full}. (c) WISE point source catalog \citep{2013yCat.2328....0C}.
\end{table}

\begin{table}
\centering
\caption{Orbital Period of TIC~184743498}
\begin{tabular}{lc}
\hline
\hline
Source & Period   \\
\hline
Gaia RV solution$^a$ & 1.053\,245\,0\,(140) \\  
ASAS-SN$^b$ BLS & 1.053\,238\,0\,(30)  \\ 
{\it TESS} BLS & 1.053\,272\,0\,(80) \\   
{\it TESS}$^c$ ETV & 1.053\,240\,0\,(3) \\ 
\hline
{\it TESS}$^c$ ETV; $\dot P_{\rm orb}/P_{\rm orb}$ [yr$^{-1}$]& $-1.03 \pm 0.02 \times 10^{-5}$ \\ 
\hline
\label{tbl:orbit}  % Table 2 of orbital periods
\end{tabular}

{\bf Notes.}  (a) Gaia SB2C orbit \citep{gaia_rvs,gaia_hot_rvs}. Gaia also reports an acceleration solution \citep{gaia_acc} which indicates non-linear proper motion.  This is consistent with the non-linear ETV behavior we find with {\it TESS}.  (b) ASAS-SN (\citealt{shappee14}; \citealt{kochanek17}). (c) See Fig.~\ref{fig:etv}. 
\end{table}

\section{Estimating the Stellar Parameters}
\label{sec:stellarparam}

We determined the stellar parameters of the stars in TIC 184743498 with an SED fitting code that utilises the $K$ velocities from Gaia (see Table \ref{tbl:mags}) and parameter ratios from the light curve modeling code {\sc Lightcurvefactory} \citep[see, e.g.][and references therein]{borkovitsetal19a,borkovitsetal20a}. 

The SED code is based on a Markov Chain Monte Carlo (MCMC; see, e.g.,\,\citealt{2005AJ....129.1706F}) fitting routine with only five adjustable parameters: (i) the primary mass, $M_1$, (ii) the secondary mass $M_2$, (iii) the mass of the tertiary star in a wide outer orbit $M_3$, (iv) the age of the system, $\tau$, and (v) the interstellar extinction, $A_V$.  The code models the following `input' information: (i) 21 SED points, (ii) the $K_1$ and $K_2$  values with uncertainties from Gaia, (iii) two ratios ($T_{\rm eff,1}/T_{\rm eff,2}$ and $(R_1+R_2)/a$) and the orbital inclination angle, $i$, from a fit to the orbital light curve with {\sc Lightcurvefactory} \citep{borkovitsetal19a,borkovitsetal20a}, and (iv) a set of stellar evolution tracks from MIST (\citealt{2016ApJS..222....8D};  \citealt{2016ApJ...823..102C};  \citealt{2011ApJS..192....3P}; \citealt{2015ApJS..220...15P};  \citealt{2019ApJS..243...10P}), where a solar composition is assumed.  We utilise the \citet{2003IAUS..210P.A20C} model stellar atmospheres for $4\,000 < T_{\rm eff} < 10\,000$\,K.  The distance is known with exquisite accuracy from Gaia parallax measurements.  

The only constraints we have on the tertiary star are (i) from the total light of the system at each of the 21 SED points, and (ii) the fact that its mass is likely to be in the range of 0.7 to 1.3 M$_\odot$ (based on the ETV curve, see Fig.~\ref{fig:etv} and the discussion of the figure).  For the latter, we simply adopted a uniform prior on $M_3$ over the range 0.5-1.6 M$_\odot$ to be conservative.

Our approach follows that of \citet{2020MNRAS.494.5118K}, \citet{rappaport21}, and \citet{rappaport22}.  The MCMC code evaluates the following five parameters: $M_1$, $M_2$, $M_3$, $A_V$, and the system age via the {\tt MIST} equivalent evolutionary phase (EEP) of the primary star. The use of EEPs as a preferred sampling parameter---rather than directly sampling the stellar age---is described in detail in \citet{2020MNRAS.494.5118K}. This procedure for the SED fitting has been utilised extensively in the study of numerous compact triples (\citealt{rappaport22}, \citealt{rappaport23}).  The validity of the results from those SED fits has been checked in a number of cases where there are several additional constraints on the system parameters. 

At each step of the MCMC routine, the procedure is as follows.  The three masses and age of the binary (via the EEP) yield the stellar radii and $T_{\rm eff}$ values through the stellar evolution tracks.  This, of course, assumes that the two EB stars have evolved in a coeval fashion and, in particular, that they have not previously exchanged any mass.  This, coupled with the model atmospheres and current value of $A_V$, allow for a model of the measured SED points. The $\chi^2$ value from the SED fit is then registered.  Additionally, the known orbital period and the two masses provide the semi-major axis of the system.  When combined with the orbital inclination angle (provided by {\sc Lightcurvefactory}), this enables us to compute $K_1$ and $K_2$, to compare with the measured Gaia values.  The $\chi^2$ values from the fit to the $K$ velocities are also recorded.  Finally, we check how well the model ratios of radii and $T_{\rm eff}$ match those provided by {\sc Lightcurvefactory}.   

\begin{figure}
\centering
\includegraphics[width=1.01\columnwidth]{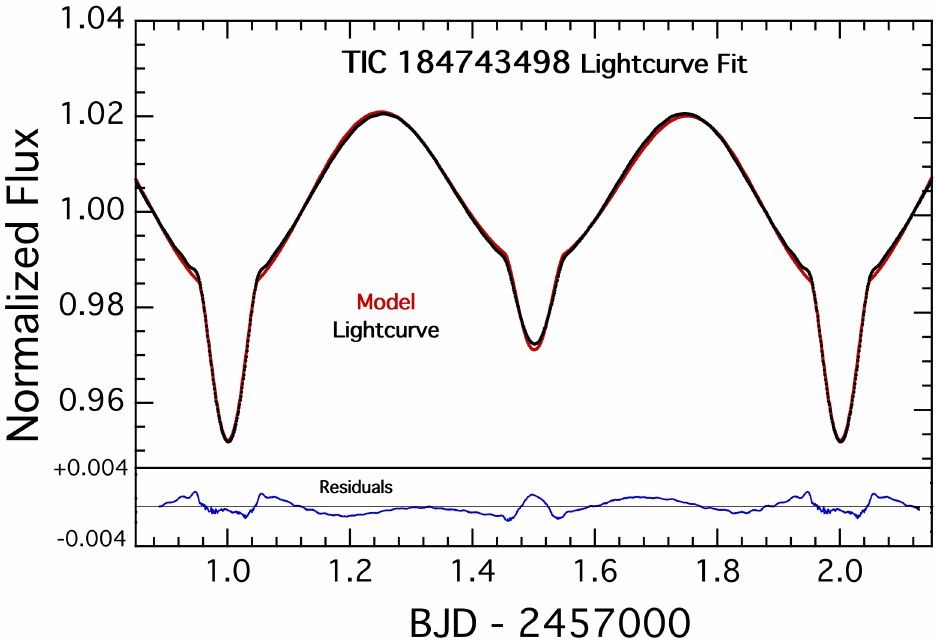}	
\caption{Model fit to the {\it TESS} sector 61 and 62 data.  The black curve is the light curve reconstructed from the first 45 orbital harmonics.  The red curve is a fitted model using {\sc Lightcurvefactory}.  See text for details and references.}
\label{fig:lightcurve_fit}
\end{figure}

The light curve that we fit with {\sc Lightcurvefactory} is shown in Figure \ref{fig:lightcurve_fit}, and is based on Sectors 61 and 62 of the {\it TESS} data.  Instead of the usual folded, or folded, binned, and averaged data,  we use a light curve that is reconstructed from the first 45 orbital harmonics\footnote{Here we used 45 harmonics as opposed to the 30 harmonics utilised for cleaning the data prior to the FT (see Sect.~\ref{sec:pulsations}).  The 45 harmonics are used here to extract the maximum high-frequency content for the light curve without running into the 51st orbital harmonic which coincides with a pulsation frequency (see Table \ref{tbl:frequencies}).} that we fit for in the {\it TESS} data using a similar technique to the one described later in Sect.\,\ref{sec:pulsations}. In principle, this synthetic light curve is the same as a folded, binned, and averaged light curve, except that (i) we avoid one of the pulsations that coincides with the 51st orbital harmonic and would otherwise appear in the folded light curve, and (ii) this also eliminates some of the high frequency noise that is not relevant to the shape of the light curve.  The {\sc Lightcurvefactory} fit to the light curve of TIC 184743498 is shown in Figure \ref{fig:lightcurve_fit} in red.  The three key parameters extracted from this fit are: $T_{\rm eff,1}/T_{\rm eff,2}= 1.241 \pm 0.010$, $(R_1+R_2)/a = 0.500 \pm 0.025$, and $i= 65^\circ \pm 0.5^\circ$.  

  % Figure 4. Model of LC

\begin{figure}
\centering
\includegraphics[width=1.0\columnwidth]{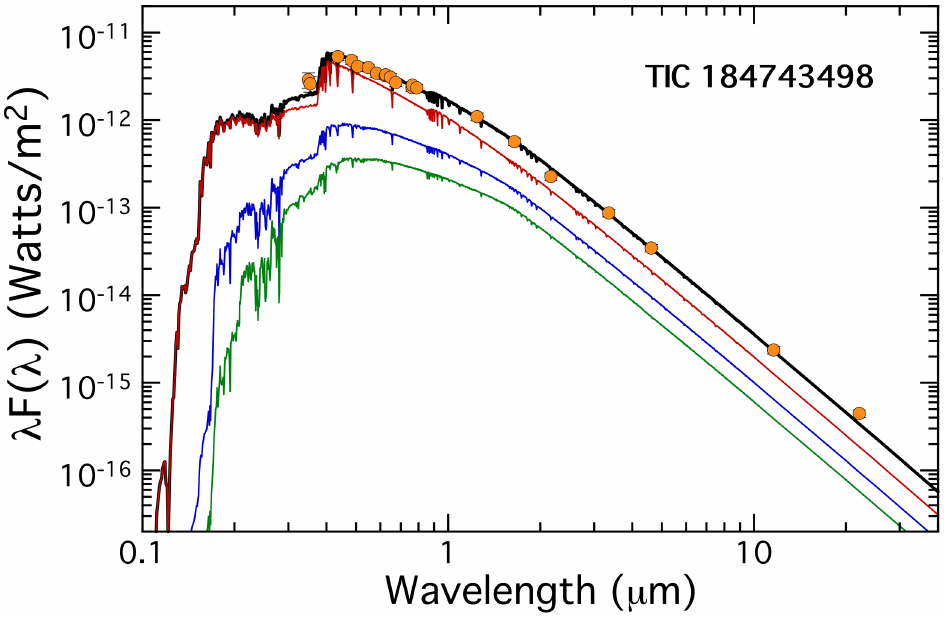}	
\caption{21 SED points for TIC 184743498 (orange points) are fitted to stellar models for the three stellar components in this system.  The contribution of the primary star is shown as a red curve, secondary is shown in blue, and tertiary is shown in green.  The total model flux is given by the black curve.  See text for the constraints that were used in the fit.}
\label{fig:SED_fit}
\end{figure}

Figure\,\ref{fig:SED_fit} shows the results of the SED fitting. We plot the measured SED points in orange, the model SED of the primary star in red, that of the secondary in blue, and of the tertiary in green. The total flux is plotted in black. We find the following stellar parameters: M$_1 = 1.83 \pm 0.07$ M$_\odot$, M$_2 = 1.37 \pm 0.046$ M$_\odot$, R$_1 = 1.72 \pm 0.06$ R$_\odot$, and R$_2 = 1.35 \pm 0.06$ R$_\odot$. The parameters of the third star in the long outer-period orbit are less constrained, but are comparable to the properties of the secondary eclipsing binary (EB) star.  The remaining fitted parameters can be found in Table \ref{tbl:parms}. 

  % Figure 5. SED fit.

\begin{figure*}
\centering
\includegraphics[width=0.8\linewidth]{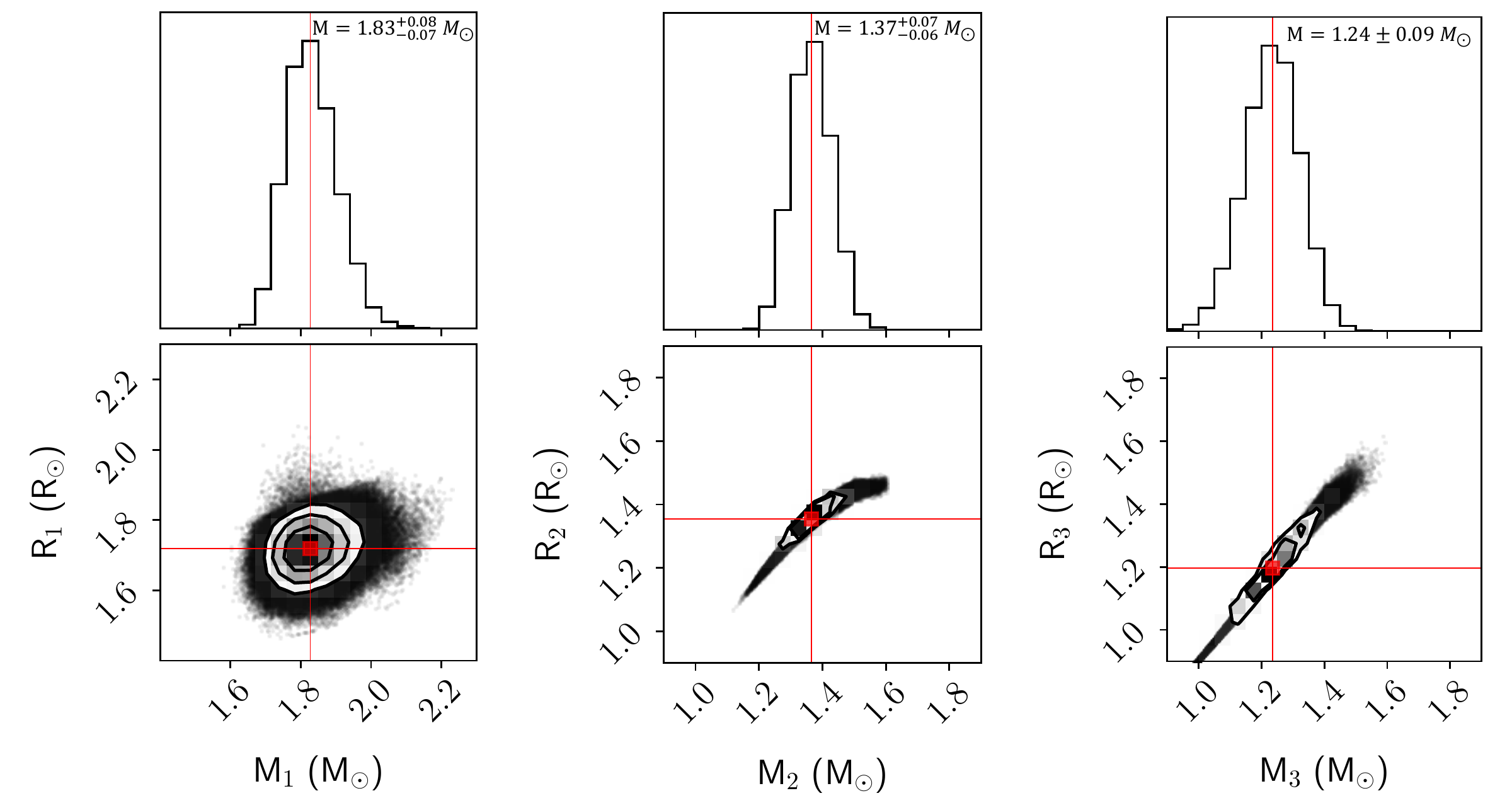}	
\caption{Correlations between the mass and radius of the three stars in the TIC 184743498 system, as well as posterior distributions of the masses.   The mass and radius of the secondary EB star and the tertiary are well correlated because these stars are still on the zero-age main sequence, where there is a one-to-one correspondence between mass and the other stellar parameters.}
\label{fig:correlations}
\end{figure*}

We show in Fig.\,\ref{fig:correlations} the correlations in the MCMC posteriors for the mass vs.~radius of the three stars in the TIC 184743498 system.  The mass and radius of the primary are fairly well localized.  By contrast, the secondary star in the EB shows a high degree of correlation between $M_2$ and $R_2$.  This results from the fact that the secondary is lower in mass and, at the age of the system, it is still firmly on the zero-age main sequence (ZAMS).  The same is true for the tertiary star (in the wide outer orbit), but the degree of correlation between $M_3$ and $R_3$ appears even more extreme due to the high degree of uncertainty in $M_3$. The top panel of Fig.\,\ref{fig:correlations} shows the 1D posterior distributions for the stellar masses of each star.

  % Figure 6. "corner" correlation plot

\begin{table}
\centering
\caption{Derived Parameters for the TIC~184743498 System}
\begin{tabular}{lc}
\hline
\hline
Input Constraints & SED + RVs$^a$ \\
\hline
Period (days) & 1.053236  \\
$K_1$ (km~s$^{-1}$)$^b$ & $119.3 \pm 2.9$\\  
$K_2$ (km~s$^{-1}$)$^b$ & $154.9 \pm 9.5$ \\
Spectral & 21 SED points$^c$  \\
Stellar evolution tracks & {\tt MIST}$^d$  \\
Distance (pc)$^b$ & $374 \pm 6$  \\
$(R_1+R_2)/a$$^e$ & $0.500 \pm 0.025$\\
$T_{\rm eff,1}/T_{\rm eff,2}$$^e$ & $1.241 \pm 0.030$\\
Stellar inclination angle$^e$ & $65.2^\circ \pm 0.5^\circ$ \\
\hline
Derived  Parameters & SED + RVs$^a$  \\
\hline
$M_1$ (M$_\odot$) &  $1.83 \pm 0.07$  \\
$M_2$ (M$_\odot$) & $1.37 \pm 0.06$  \\
$M_3$ (M$_\odot$) & $1.23 \pm 0.09$ \\
$R_1$ (R$_\odot$) & $1.72 \pm 0.06$ \\
$R_2$ (R$_\odot$) & $1.35 \pm 0.06$ \\
$R_3$ (R$_\odot$) & $1.19 \pm 0.11$ \\
$T_{\rm eff,1}$ (K) & $8500 \pm 300$ \\
$T_{\rm eff,2}$ (K) & $6870 \pm 200$ \\
$T_{\rm eff,3}$ (K) & $6450 \pm 270$ \\
$L_{\rm 1,bol}$ (L$_\odot$) & 14.1 $\pm 2.2 $ \\
$L_{\rm 2,bol}$ (L$_\odot$) & $3.7 \pm 0.8$ \\
$L_{\rm 3,bol}$ (L$_\odot$) & $2.3 \pm 0.8 $ \\
%$i$ (deg)$^e$ & $65 \pm 0.5$$^f$   \\
$a$ (R$_\odot$) & $6.41 \pm 0.09$ \\
$R_1/R_L$ & $0.66 \pm 0.02$ \\
age (Myr) & $460 \pm 120$ \\
$A_V$ & $0.35 \pm 0.10$  \\
\hline

\label{tbl:parms} 
\end{tabular}   % Table 3. parameter table

{\bf Notes.} (a) MCMC fits to the measured RV amplitude plus the SED points.  The assumption is made that the two stars are coeval in their evolution, and have not exchanged any mass. (b) Gaia \citep{gaia_rvs} (c) VizieR: \citep{ochsenbein00}; A.-C. Simon \& T. Boch: http://vizier.unistra.fr/vizier/sed/). (d) MIST (\citealt{2016ApJS..222....8D};  \citealt{2016ApJ...823..102C}. (e) {\sc Lightcurvefactory} \citep{borkovitsetal19a,borkovitsetal20a} fit to the {\it TESS} orbital light curve. 

\end{table}

\section{Pulsations} 
\label{sec:pulsations}

We analyzed the pulsations of TIC 184743498 by applying a discrete Fourier transform program to the raw {\it TESS} data.  We performed FTs following various stages of cleaning: We first removed the orbital harmonics from the data, and then removed the more prominent pulsation mode frequencies. As mentioned in Sect.~\ref{subsec:tessdata}, we use Sectors 61 and 62 for pulsation analysis. 

We started by subtracting the first 30 orbital harmonics from the data, the highest frequency of which is near 28.5 d$^{-1}$.  This was done via a simultaneous linear least-squares fit to 30 sines and 30 cosines to represent the orbital modulations.  We purposely stopped the cleaning at 30 orbital harmonics so as not to inadvertently remove any natural pulsation frequencies that happen to lie near an orbital harmonic or are, in fact, tidally excited at exactly that harmonic.  The amplitudes of the orbital harmonics above this frequency are small enough so that they are not likely to be confused with stellar pulsations. Furthermore, possible residual orbital harmonics will have a phase in an echelle diagram (see below) close to 0 or 1, and will therefore be easily recognized. The middle panel of Fig.\,\ref{fig:ft1} shows the resultant FT in the frequency range from 0 to 80 d$^{-1}$ after 30 orbital harmonics have been removed. We see a rich spectrum of pulsations within the range of 38-56\,d$^{-1}$.  The bottom panel of Fig.\,\ref{fig:ft1} shows the transform after removing the 13 most prominent frequencies (enumerated in Table \ref{tbl:frequencies}).

To better visualize the organization of the pulsations in TIC~184743498, we generated an echelle diagram. Such a diagram plots the frequency of a pulsation on the y-axis against the echelle phase on the x-axis; this so-called ``echelle phase'' is the pulsation frequency modulo the orbital frequency, normalized by the orbital frequency. In the creation of an echelle diagram, we set a threshold for the minimum Fourier amplitude ($\gtrsim 7\sigma$), such that only highly significant peaks are represented. 

\begin{figure*}
\centering
\includegraphics[width=0.65\linewidth,angle=0]{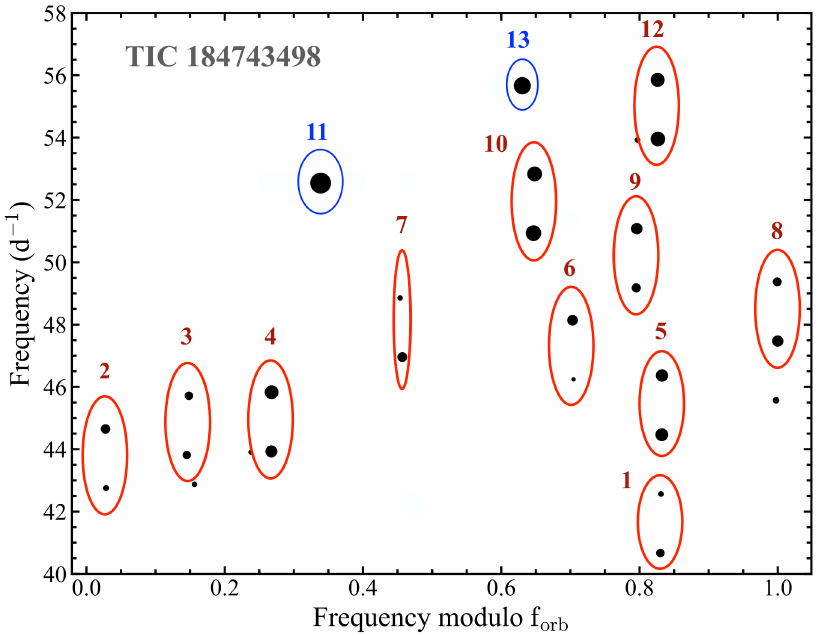} 	
\caption{Echelle diagram following a cleaning of the data that removed the first 30 orbital harmonics. This diagram shows pulsation peaks as a function of their frequency and echelle phase, i.e., the normalized pulsation frequency modulo the orbital frequency. Points are sized linearly by their Fourier amplitude.  Red ellipses mark the pulsation multiplets that we have identified.  These are numbered according to increasing central frequency to match the notation in Table \ref{tbl:frequencies}. }
\label{fig:echelle}
\end{figure*}

Fig. \ref{fig:echelle} displays the echelle diagram for TIC~184743498 following the removal of 30 orbital harmonics. The amplitude threshold is 0.05 ppt ($\sim$$7\,\sigma$). We limit the echelle diagram to the relevant frequency range where the vast majority of peaks are located, and size each dot by linearly scaling the Fourier amplitude to emphasize the more prominent peaks. We see 11 doublets (also referred to as ``multiplets''), which are a vertically aligned pair of filled circles separated by twice the orbital frequency. These doublets are encircled in red ellipses and labeled by increasing frequency. These results are also tabulated in Table \ref{tbl:frequencies}. We encircle two singlets in blue ellipses. 

For each multiplet, we find the frequencies of the observed peaks (marked by ellipses in the echelle diagram) and then calculate the inferred (unseen) central frequency at which the star is actually pulsating. All the measured and inferred frequencies of the multiplets and singlets are summarized in Table~\ref{tbl:frequencies}.  

  % Figure 7.  labelled echelle

\begin{table*}
\centering
\caption{Dominant Pulsation Frequencies in the TIC~184743498 System}
\begin{tabular}{lcccc}
\hline
\hline
Mode Name  &  Frequency$^a$ &  Amplitude &  Phase$^b$ &  Echelle Phase   \\
                    & d$^{-1}$  & ppt  & radians          &        cycles      \\
\hline
Uncertainty: & 0.0006 & 0.007  & $\sim$0.03 & $\sim$0.002  \\
\hline
$\nu_1-\nu_{\rm orb}$ & 40.6653   &	0.234    &  $-1.10$ &  0.830      \\    %0.231
$\nu_1$              &             (41.6152)  &                &   &                 \\            %0.023   dphi = +0.13   Y10x
$\nu_1+\nu_{\rm orb}$ & 42.5651   &   0.143   & $-1.23$ & 0.831     \\     %0.141
\hline
$\nu_2-\nu_{\rm orb}$ & 42.7525    & 0.149  &  $-0.60$  &   0.028     \\   % 0.149
$\nu_2$                           & (43.7016)   &              &   &                   \\         % 0.020   dphi = -0.06   Y10x
$\nu_2+\nu_{\rm orb}$ & 44.6507   &  0.274  &  $-0.54$  &  0.028    \\    % 0.248
\hline
$\nu_3-\nu_{\rm orb}$ & 43.8129   &	0.225  & $+1.45$ &	0.145     \\       % 0.188 **
$\nu_3$                       & (44.7638)$^c$ &                 &   &    \\                   % 0.058   dphi = -0.26   Y10x
$\nu_3+\nu_{\rm orb}$ & 45.7147  &	0.225  & $+1.71$  &   0.148    \\     % 0.0.222
\hline
$\nu_4-\nu_{\rm orb}$ & 43.9289	& 0.399     &  $-2.20$  & 	0.267   \\  % 0.385
$\nu_4$                       & (44.8785) &             &    &                \\                   % 0.025  dphi = -0.01   Y10x
$\nu_4+\nu_{\rm orb}$ & 45.8282	& 0.507 & $-2.19$  &  0.268    \\      % 0.505      
\hline
$\nu_5-\nu_{\rm orb}$ & 44.4652	& 0.449  & $-2.51$ &  0.832   \\      % 0.443
$\nu_5$                       &  (45.4147)            &                & &                 \\      % 0.018   dphi = -3.12   Y10y
$\nu_5+\nu_{\rm orb}$ & 46.3643	 & 0.419  & $+0.61$ &  0.832    \\     % 0.420
\hline
$\nu_6-\nu_{\rm orb}$ & 46.2429	& 0.118   & $-0.55$ &    0.705   \\    % 0.107
$\nu_6$                       & (47.1917)  &             &  &                 \\                   % 0.009   dphi = -3.13  (Y10y)
$\nu_6+\nu_{\rm orb}$ & 48.1405	& 0.323  & $+2.58$ &    0.703    \\   % 0.329   
\hline
$\nu_7-\nu_{\rm orb}$ & 46.9572	& 0.288  & $-2.05$ &   0.457   \\      %  0.278
$\nu_7$                       & (47.9047)               &              &  &                   \\    %  0.022   dphi = -0.41  (Y10x?)
$\nu_7+\nu_{\rm orb}$ & 48.8532	& 0.137  &  $-1.64$ &   0.454    \\     %  0.169	
\hline
$\nu_8-\nu_{\rm orb}$ & 47.4729	& 0.362  & $-2.00$ &   1.000    \\      % 0.354
$\nu_8$                       & (48.4219)$^d$  &           &  &                 \\               % 0.022   dphi = -3.36   Y10y
$\nu_8+\nu_{\rm orb}$ & 49.3710	& 0.247  & $+1.36$ &   0.999    \\      % 0.239
\hline
$\nu_{9}-\nu_{\rm orb}$ & 49.1773  &  0.259	& $-2.05$ & 0.795     \\      % 0.261    
$\nu_{9}$                       & (50.1271)    &                  &  &              \\               % 0.042   dphi = +0.01   Y10x
$\nu_{9}+\nu_{\rm orb}$ & 51.0769  &  0.368 & $-2.06$ & 	0.796     \\   % 0.348    
\hline
$\nu_{10} -\nu_{\rm orb}$ & 50.9351  &	0.630  & $+0.63$ &  0.647   \\     % 0.600
$\nu_{10}$                      &  (51.8853)  &               &  &                \\                 % 0.006   dphi = +3.16   Y10y
$\nu_{10}+\nu_{\rm orb}$ & 52.8356  &	0.579 & $-2.53$ & 	0.648    \\     %	0.583
\hline
$\nu_{11}$                        & 52.5418   &  1.113	&  ...  &     0.339     \\             % 1.096
\hline
$\nu_{12}-\nu_{\rm orb}$ & 53.9543  &   0.563  & $+1.42$ & 0.826    \\         % 0.557
$\nu_{12}$                       & (54.9036)            &   &                &                \\       % 0.013   dphi = +3.15   Y10y
$\nu_{12}+\nu_{\rm orb}$ & 55.8529  &   0.513 & $-1.73$ & 0.827    \\           % 0.505
\hline
$\nu_{13}$                       & 55.6671   & 0.765	& ... & 0.631	  \\                   % 0.762; upper ccomponent: 0.095
\hline
\label{tbl:frequencies} 
\end{tabular}   % Table 4 of frequencies.

{Notes. (a) A least-squares fit of the 13 most prominent pulsation modes.  Numbers in parentheses are the unseen central frequencies of the doublets, and hence are the actual mode frequencies.  (b) The zero point for the phases has been chosen to be the time of the primary eclipse: $t_0 = {\rm BJD}\,2459988.9789$.  For later reference in the paper, we note that phase differences between the two components with values near zero correspond to what we call a $Y_{10,x}$ mode, while those with phase differences closer to $\pi$ are called $Y_{10,y}$ modes (see Eqns.~(\ref{eq:xi+}) and (\ref{eq:xi-})). (c) This central frequency was likely directly observed at $\nu = 44.7594$ d$^{-1}$, amplitude = 0.073 ppt, and echelle phase 0.142.  (d) This has an echelle phase that is completely consistent with an orbital harmonic, but its amplitude is too high for that to be the case.  Thus, this is either a natural frequency that just happens to coincide with an orbital harmonic to within the uncertainties (0.4\% random chance), or is a mode driven by the orbit. Moreover, the echelle diagram (Fig.~\ref{fig:echelle}) appears to show a third component at $\nu=45.5715$ d$^{-1}$, but we have shown from the use of all the data, including Sectors 34 and 35, that this is not part of the same doublet.} 
\end{table*}

\section{Reconstructing the Pulsation Amplitudes and Phases}
\label{sec:reconstruct}

The amplitudes and phases of the elements of each of the 11 pulsation doublets (see Fig.~\ref{fig:echelle}) carry all the information about a particular pulsation mode that can be extracted from the {\it TESS} data.  However, the specific pulsation modes cannot be entirely inferred merely by inspecting the echelle diagram.  In this regard, it is extremely useful to reconstruct the behavior of the pulsation amplitudes and phases as a function of orbital phase. To do this, we analytically reconstruct the amplitudes and phases of 9 of the doublets in the echelle diagram (all except for doublets 6 and 7, whose phase and amplitude behavior are not robustly characterized).  We follow the formalism for reconstructing multiplets provided in Eqns.~(2)-(6) of \citet{jayaraman22}.  

To be conservative, we force fit a quintuplet of components to each doublet, at frequencies $\nu_0, \nu_0 \pm \nu_{\rm orb}$ and $\nu_0 \pm 2 \nu_{\rm orb}$.   If at one or more of the 5 frequencies there is no significant FT amplitude, then the only contribution to the reconstruction will be a small amount of added noise. However, there may be a weak (albeit real) signal there which did not surpass the echelle threshold, and we cannot neglect its contribution.  Note that we use the same reconstruction process for all the pulsation modes, regardless of spacing or status as a singlet or doublet. Finally, we chose the same reference eclipse time, $t_0$, as was used in the pulsation analysis in Section \ref{sec:pulsations}. 
 
\begin{figure*}
\centering
\includegraphics[width=0.82\linewidth,angle=0]{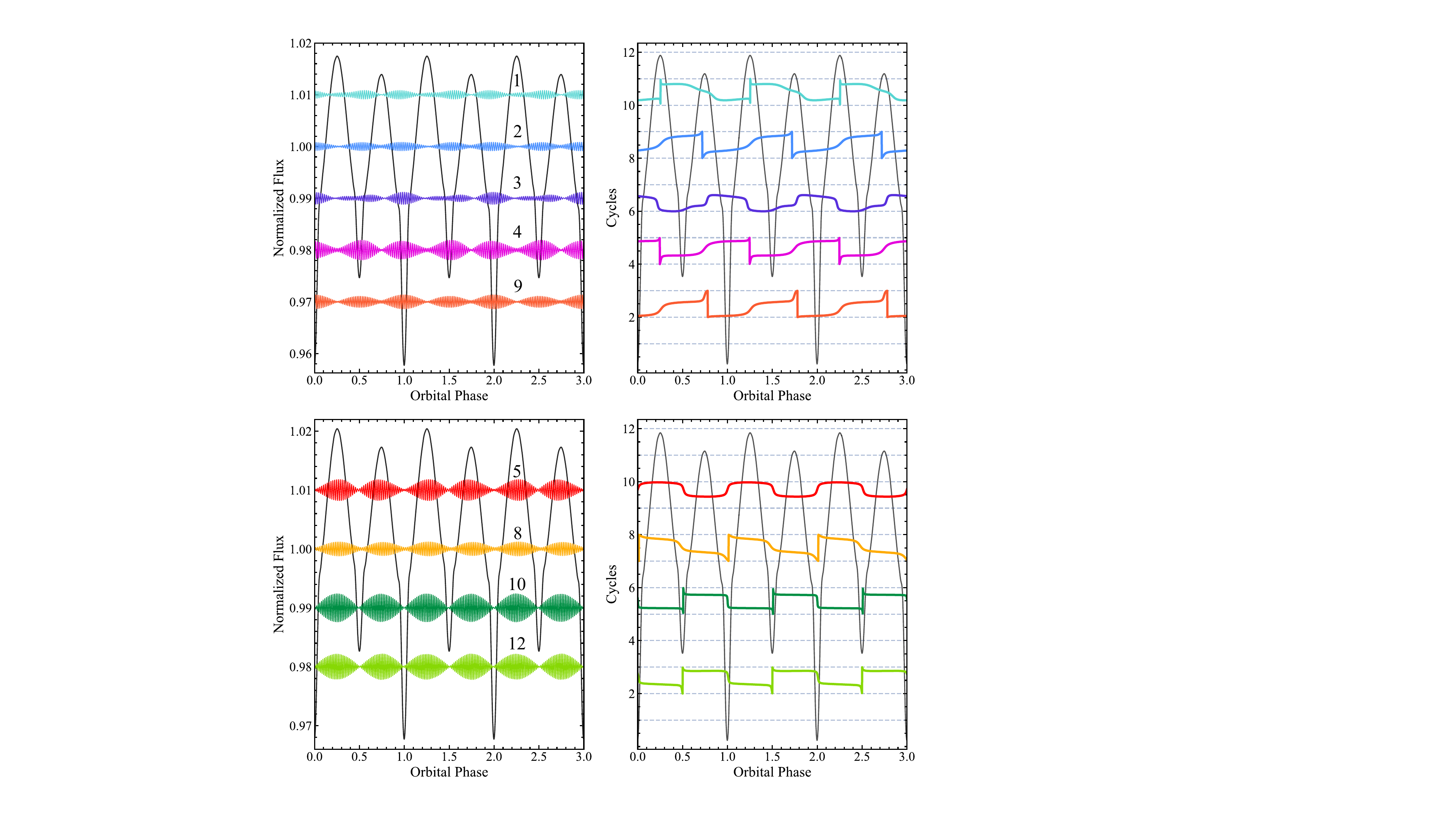} 	
\caption{Left and right panels: The reconstructed pulsation amplitude and phase variations as a function of orbital phase for the five doublets that we have classified as $Y_{10,x}$ pulsation modes with a pulsation axis along the tidal axis ($\nu_{1}$ through $\nu_{4}$ and $\nu_{9}$); the frequencies are listed in Table \ref{tbl:frequencies}. The reconstructed pulsations have been vertically offset from each other by 0.01 for clarity, and each pulsation amplitude has been doubled from its actual value to make it more visible.  The phase reconstructions are offset vertically from one another by 2 cycles. We label each reconstructed mode with its frequency numbering.  The black curve superposed on the plots is the reconstructed orbital modulation.  See text for details of the reconstructions.} 
\label{fig:reconstruction1}
\end{figure*}

\begin{figure*}
\centering
\includegraphics[width=0.82\linewidth]{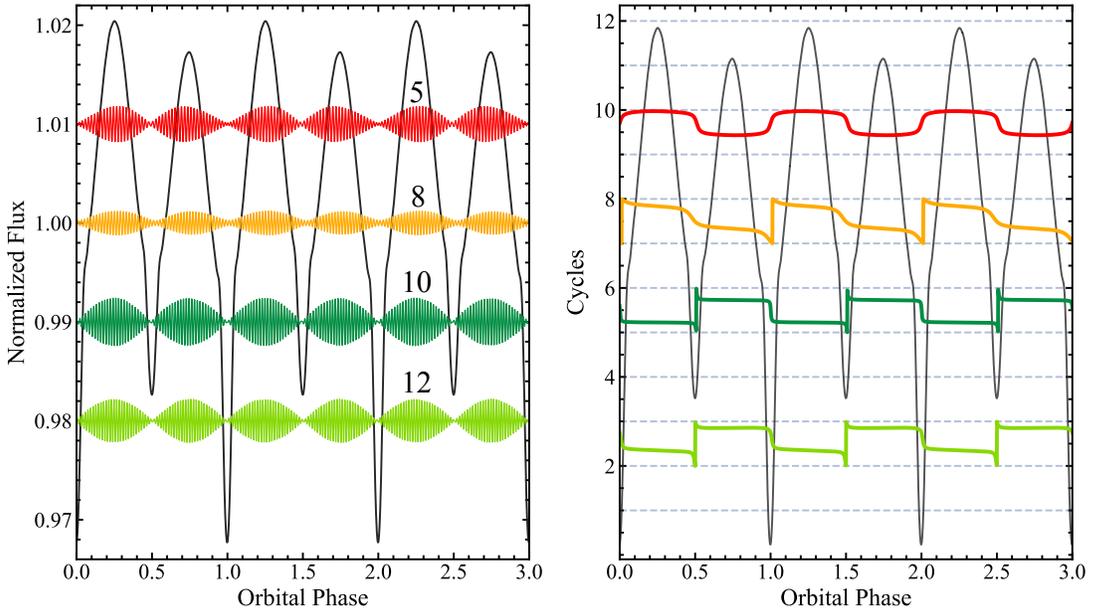}	
\caption{Left and right panels: The reconstructed pulsation amplitude and phase variations as a function of orbital phase for the four doublets that we have classified as $Y_{10,y}$ pulsation modes with a pulsation axis along the $y$ direction (perpendicular to the tidal and angular momentum axes).  The frequencies $\nu_{5}$, $\nu_{8}$, $\nu_{10}$, and $\nu_{12}$ are listed in Table \ref{tbl:frequencies}. The reconstructed pulsations have been vertically offset from each other by 0.01 for clarity, and each pulsation amplitude has been doubled from its actual value to make it more visible.  The phase reconstructions are offset vertically from one another by 2 cycles. We label each reconstructed mode with its frequency numbering.  The black curve superposed on the plots is the reconstructed orbital modulation.  See text for details of the reconstructions.} 
\label{fig:reconstruction2}
\end{figure*}

Fig.\,\ref{fig:reconstruction1} shows the amplitude-phase plots for the central frequencies $\nu_1$ though $\nu_4$, and $\nu_9$, in Table \ref{tbl:frequencies}, and Fig.~\ref{fig:reconstruction2} gives the amplitude-phase plots for frequencies $\nu_5$, $\nu_8$, $\nu_{10}$, and $\nu_{12}$. The amplitude reconstructions are shown in the left panels, while the phase variation over the orbit is displayed in the right panels.  During the reconstruction, we scaled the pulsation amplitude up by a factor of 2 for better visibility. The two different figures represent different kinds of pulsation modes.

We have collected in Fig.\,\ref{fig:reconstruction1} all five doublet pulsation modes whose amplitude maxima occur at the eclipses and have $\pi$ phase jumps near the ellipsoidal light variation (ELV) peaks. By contrast, Fig.\,\ref{fig:reconstruction2} displays all four doublets whose amplitude maxima occur at the ELV peaks and have $\pi$ phase jumps near the eclipses.  The pulsation modes in Fig.\,\ref{fig:reconstruction1} can all be described by dipole modes of the form $Y_{10}$, namely with $\ell =1$ and $m = 0$, if the pulsation axis has been tidally tilted into the orbital plane and follows the tidal axis as it orbits with the binary.  At low inclination angles, even lower than the $65^\circ$ for this system, the doublet nature of such a pulsation mode along the tidal axis will be preserved (see, e.g., \citealt{reed05}). This means no central peak will appear in the echelle diagram, and indeed this is the case as we look at doublets for $\nu_1$, $\nu_2$, $\nu_3$, $\nu_4$, and $\nu_9$ in Fig.\,\ref{fig:echelle}.  We hereafter refer to these as $Y_{10,x}$ modes.

  % Figure 8 1-5 pulsation + amplitudes

By contrast, the reconstructions in Fig.\,\ref{fig:reconstruction2} all have pulsation maxima and $\pi$ phase shifts at orbital phases that are $90^\circ$ displaced from those in Fig.~\ref{fig:reconstruction1}. Thus, one might be tempted to explain these pulsations as dipole modes of the form $Y_{11}$ ($\ell =1$, and $|m| = 1$), again for a tidally tilted pulsation axis.  However, for this system's relatively low orbital inclination angle, strong central frequency components must appear in the echelle diagram for $Y_{11}$ modes, yet we do not see such a central component for any of the doublets in Fig.\,\ref{fig:echelle}. This is equivalent to noting that a $Y_{11}$ mode tilted along the tidal axis which, in turn, is at an inclination of 65$^\circ$, could not have zero pulsation amplitude at the eclipses -- as we see in Fig.\,\ref{fig:reconstruction2}.

\begin{figure}
\centering
\includegraphics[width=1.0\linewidth]{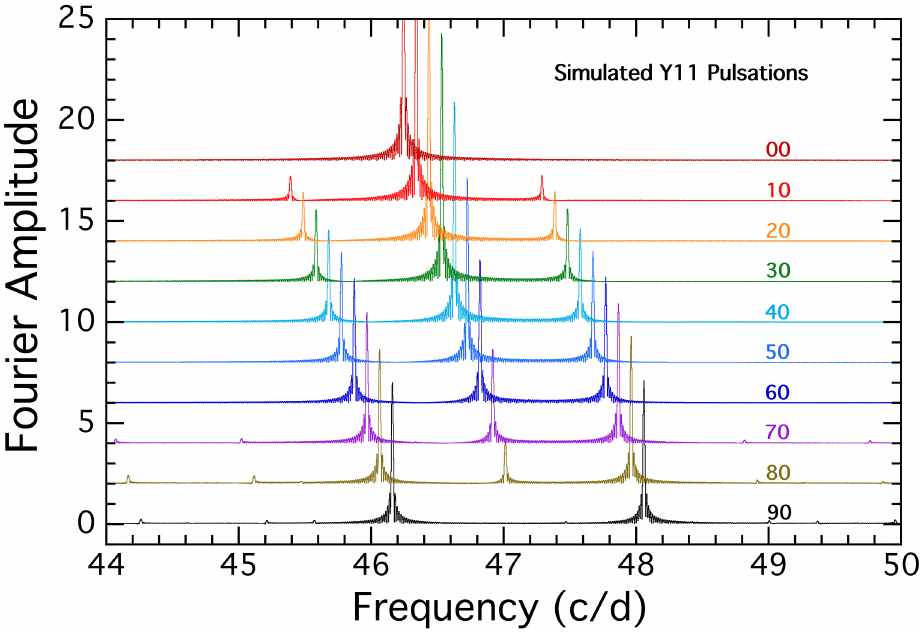}	
\caption{Fourier transforms of simulated pulsation for a mode $Y_{11}$ with a pulsation axis along the tidal axis of the binary (x-axis). The orbital inclination angle for each simulation is labeled in color.  The curves for different inclinations are separated both vertically and horizontally for clarity.  A central peak starts to appear at $i  \lesssim 80^\circ$, and is nearly equal to the split sidelobes by $i=60^\circ$.}  
\label{fig:simft}
\end{figure}

In Fig.~\ref{fig:simft}, we show explicitly how the FT of a tidally tilted $Y_{11}$ mode varies with orbital inclination angle. As is clear from the figure, a significant central peak begins to appear at $i \lesssim 80^\circ$, and is nearly equal to the split sidelobes by $i=60^\circ$.  Thus, it would be implausible to have such a mode produce the types of doublets we see in Fig.\,\ref{fig:echelle}, which have neither a strong central peak nor non-trivial pulsations at the eclipses.

  % Figure 9. pulsations/phases of multiplets 6-9.

  % Figure 10. FTs for simulated modes.

We therefore propose that the four modes shown in Fig.\,\ref{fig:reconstruction2} are actually $Y_{10,y}$ modes with the pulsation axis along the ``$y$-axis'', which is perpendicular to the tidal axis ($x$) and the orbital angular momentum axis ($z$).  This hypothesis resolves the issue of the missing central component of the doublets, and zero pulsation amplitude at the eclipses, while being fully consistent  with the pulsation amplitudes and phase shifts with respect to orbital phase. Thus, this represents the first discovery of a stellar pulsator with a pulsation axis along the $y$ direction.  

In the same spirit as the above hypothesis, we also propose that the two singlets found in this source and listed in Table~\ref{tbl:frequencies} are $Y_{\rm 10}$ modes with a pulsation axis along $z$.  Given that their two frequencies are 52.5411 and 55.6664 d$^{-1}$, with a difference of $\sim$ 3 d$^{-1}$, the interval here is too narrow to host radial modes with such frequencies. This hypothesis is further explained in Section \ref{sec:radial_modes}.

Finally, we note that the modes with frequencies $\nu_6$ and $\nu_7$, which do not appear in Figs.~\ref{fig:reconstruction1} and \ref{fig:reconstruction2}, each have markedly different amplitudes for their two doublet peaks.  However, the reconstructed pulsation amplitude and phase behavior with orbital phase of $\nu_6$ very closely resembles those in Fig.~\ref{fig:reconstruction2}, i.e., $Y_{10,y}$ modes, while $\nu_7$ would be associated with those modes in Fig.~\ref{fig:reconstruction1}, i.e., $Y_{10,x}$.  Thus, these two doublets are likely distorted $Y_{10}$ tidally tilted modes.

\section{Density of Radial Models}
\label{sec:radial_modes}

In Sect.\,\ref{sec:pulsations}, we reported two singlets spaced in frequency by only $\sim$3 d$^{-1}$. In the presence of tidally tilted pulsations, such singlets are most readily explained by radial modes, as their pulsational amplitudes and phases will not be modulated over the orbit. We investigate whether radial modes are a plausible interpretation for those two frequencies.

\begin{figure}
\centering
\includegraphics[width=1.0\columnwidth]{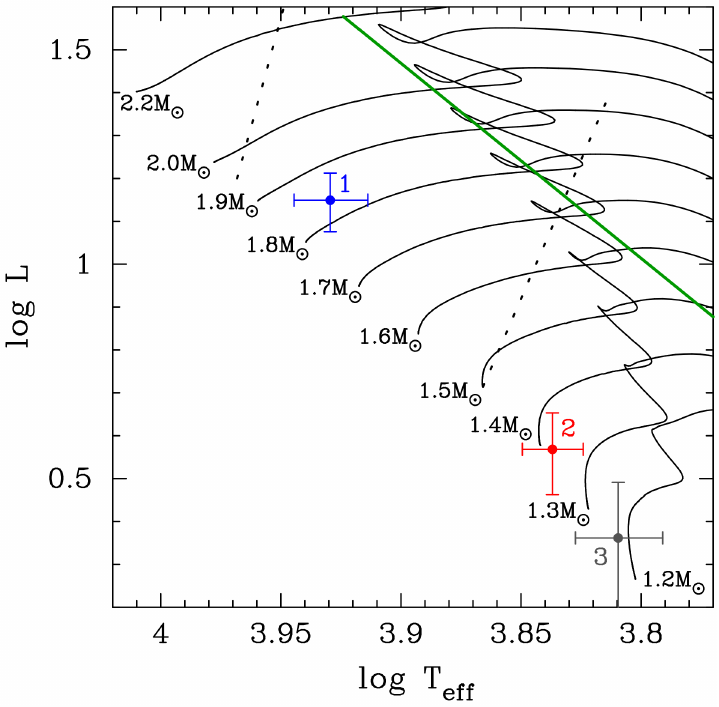}
\caption{Theoretical HRD with the locations of the three components of TIC 184743498 (stars 1, 2, and 3) indicated. The evolutionary tracks are the same as were used in the SED fit shown in Fig.~\ref{fig:SED_fit}. The dashed black lines are the observed boundaries of the $\delta$ Scuti star instability strip \citep{2019MNRAS.485.2380M}. The green line connects models that best represent the observed pulsation frequencies $\nu_{11}$ and $\nu_{13}$ with radial modes. }
\label{fig:hrd}
\end{figure}

To this end, we computed single-star pulsational models using the latest version of the Warsaw-New Jersey stellar evolution and pulsation code \citep[e.g.,][]{1998A&A...333..141P}, for a solar chemical composition \citep{2004A&A...417..751A} and a rotational velocity of 100 km\,s$^{-1}$ at the ZAMS. We evaluated the frequency differences of consecutive radial overtones of modes in the domain of 50 -- 60\,d$^{-1}$ along each evolutionary track and determined the locations of models that would reproduce the observed frequency difference $\nu_{13}-\nu_{11}$ at the radial mode frequency closest to $\nu_{13}$. Figure\,\ref{fig:hrd} shows the loci of those models, as well as of the stellar components of the TIC 184743498 system, in a theoretical Hertzsprung-Russell Diagram (HRD).

  % Figure 11. theoretical HRD.

Inspection of Fig.\,\ref{fig:hrd} clearly demonstrates that (i) only the primary star (star 1) is expected to exhibit $\delta$ Scuti type pulsations, and (ii) frequencies $\nu_{11}$ and $\nu_{13}$ cannot both be radial modes, as all stellar models that produce consecutive radial modes that are closely spaced in frequency are too evolved. Models producing $\nu_{11}$ and $\nu_{13}$ as non-consecutive radial modes would be even more evolved.

\begin{figure}
\centering
\includegraphics[width=1.0\columnwidth]{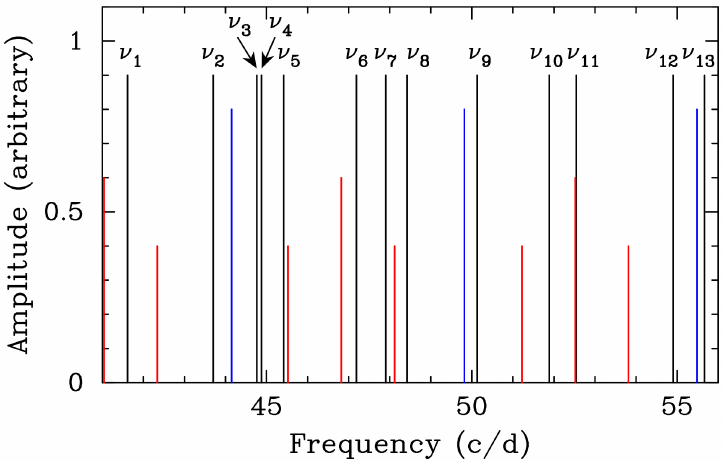}	
\caption{Schematic pulsation frequency spectrum of TIC 184743498. The black lines denote the observed centroid mode frequencies, whereas the blue lines indicate the theoretical model radial-mode frequencies. Red lines represent theoretical dipole mode frequencies, with those of `amplitude' 0.6 corresponding to $m = 0$ and those of `amplitude' 0.4 to $|m| = 1$.}
\label{fig:dipolemodes}
\end{figure}

Additionally, we look at the expected pulsation spectrum of dipole modes of a single star with the parameters we derived for the primary star in the TIC 184743498 system. The corresponding pulsation frequencies were computed analogously to those of the radial modes with the Warsaw-New Jersey code, for a model with log $T_{\rm eff}=3.9206$, and log $L=1.1302$ (cf. Table\,\ref{tbl:parms}). Nonradial mode rotational frequency splittings were computed to first order in stellar rotation frequency. The schematic frequency spectrum so derived is shown in Fig.\,\ref{fig:dipolemodes}.

  % Figure 12. schematic spectrum

Evidently, the expected spectrum of axisymmetric dipole modes ($Y_{10}$) is not dense enough to explain all the observed frequencies.  As the underlying model is fairly unevolved, no avoided crossings (mixed modes) are present in the theoretical frequency spectrum; the modes shown here are of radial overtone 5 to 7. However, by including sectoral dipole modes ($Y_{1\pm 1}$), the density of the observed apparent dipole frequency spectrum can be explained.

The above arguments lead us to the conclusion that at least one of the $\nu_{11}$ and $\nu_{13}$ singlets, and possibly both, must indeed be a $Y_{10}$ mode with a pulsation axis along $z$, i.e., $Y_{10,z}$ modes.

\section{Tidal Mode-Coupling Models}
\label{sec:models}

As suggested in Section \ref{sec:pulsations}, the pulsations of TIC 184743498 can be best explained by three different pulsation axes. In particular, it appears that there are five dipole $Y_{10}$ modes with pulsation axis aligned with the tidal (or $x$ axis), while four are dipole $Y_{10}$ modes with a pulsation axis aligned along the $y$ axis (i.e., in the orbital plane, but orthogonal to the $x$ axis).  Here we justify this hypothesis with a straightforward physical argument. 

Rather than assuming that tidal distortion aligns pulsations with the tidal axis, we show that tidal coupling between modes naturally produces modes with three different pulsation axes. Consider an $\ell=1$ triplet, now using the rotation/orbital axis as a reference axis. We formally define the $x-$axis in the direction of the companion, the $z-$axis in the direction of the spin/orbital angular momentum, with the $y-$axis lying in the orbital plane. The tidal distortion is dominated by the $\ell=m=2$ component of the tidal potential, which couples the $\ell=m=1$ mode with the $\ell=1$, $m=-1$ mode. Writing out the coupled eigensystem for this mode triplet (i.e., Eqn.~8 of \citealt{2020MNRAS.tmp.2716F}), we find
\begin{align}
\label{eq:matrixcoup}
&
\begin{bmatrix}
\omega_\alpha^2 + \delta \omega^2_0 & 0 & 0 \\
0 & \omega_\alpha^2 + \delta \omega^2_1 & \delta V_{1-1} \\
0 & \delta V_{1-1} & \omega_\alpha^2 + \delta \omega^2_{-1}
\end{bmatrix}
\begin{bmatrix}
a_0 \\
a_1  \\
a_{-1} 
\end{bmatrix}
\newline \nonumber \\
&= \omega^2
\begin{bmatrix}
1 & 0 & 0 \\
0 & 1 & \delta T_{1-1} \\
0 & \delta T_{1-1} & 1 
\end{bmatrix}
\begin{bmatrix}
a_0 \\
a_1  \\
a_{-1} 
\end{bmatrix}
\end{align}
Here, $\omega$ is the frequency of a mode of the coupled system, $\omega_\alpha^2$ is the unperturbed frequency of the mode triplet, and $\delta V_{1-1}$ and $\delta T_{1-1}$ are tidal coupling coefficients. The values of ${\bf a}$ are the relative contributions of the $Y_{1m}$ components of the $\ell=1$ triplet to the new eigenfunctions of the coupled triplet. Solving the eigensystem of Eqn.~\ref{eq:matrixcoup} yields the three mode frequencies $\omega^2$ and eigenfunctions ${\bf a}$ of the perturbed mode triplet. 

Since the $m=0$ mode remains uncoupled in this simple scenario, its eigenfrequency is slightly perturbed and its eigenfunction remains unchanged. It is an $m=0$ mode about the $z-$axis, i.e., the spin/orbital axis. The $m=\pm1$ modes are coupled, however, and the eigensystem of Eqn.~\ref{eq:matrixcoup} for those modes reduces to
\begin{align}
\label{eq:matrixcoup2}
&
\begin{bmatrix}
1 & -\delta T_{1-1} \\
-\delta T_{1-1} & 1 
\end{bmatrix}
\begin{bmatrix}
\omega_\alpha^2 + \delta \omega^2_1 & \delta V_{1-1} \\
\delta V_{1-1} & \omega_\alpha^2 + \delta \omega^2_{-1}
\end{bmatrix}
\begin{bmatrix}
a_1  \\
a_{-1} 
\end{bmatrix}
\newline \\
&\simeq
\begin{bmatrix}
\omega_\alpha^2 + \delta \omega^2_1 & \delta \omega_{\rm tide}^2 \\
\delta \omega_{\rm tide}^2 & \omega_\alpha^2 + \delta \omega^2_{-1}
\end{bmatrix}
\begin{bmatrix}
a_1  \\
a_{-1} 
\end{bmatrix}
\newline \\
&\simeq \omega^2
\begin{bmatrix}
a_1  \\
a_{-1} 
\end{bmatrix} \, ,
\end{align}
where
\begin{equation}
    \delta \omega_{\rm tide}^2 = \delta V_{1-1} - \omega_\alpha^2 \delta T_{1-1} \, .
\end{equation}
Here, we have assumed small perturbations such that $\delta \omega_1^2$, $\delta \omega_{-1}^2$, $\delta V_{1-1} \ll \omega_\alpha^2$, and $\delta T_{1-1} \ll 1$, and we have dropped second-order terms in these small quantities. 

In the limit that $\delta \omega_{\rm tide}^2 \ll \delta \omega_1^2 - \delta \omega_{-1}^2$, then this eigensystem reduces to the uncoupled system, with frequencies $\omega^2 = \omega_\alpha^2 + \delta \omega_1^2$ and $\omega^2 = \omega_\alpha^2 + \delta \omega_{-1}^2$, and eigenfunctions of $Y_{11}$ and $Y_{1-1}$. All modes have the $z-$axis as their pulsation axis.

However, in the limit of strong tidal coupling such that $\delta \omega_{\rm tide}^2 \gg \delta \omega_1^2- \delta \omega_{-1}^2$, solving the eigensystem yields the new mode eigenfrequencies
\begin{equation}
    \omega_{\pm}^2 = \omega_\alpha^2 \pm \delta \omega_{\rm tide}^2 \, .
\end{equation}
with corresponding eigenvectors 
\begin{equation}
{\bf a}_\pm =
\begin{bmatrix}
0 \\
1  \\
\pm 1 
\end{bmatrix} 
\end{equation}
Hence, the new mode eigenfunctions are equal superpositions of $Y_{1,1}$ and $Y_{1,-1}$, with angular flux perturbation patterns
\begin{equation}
    \delta F_\pm \propto \big[ Y_{11}(\theta,\phi) \pm Y_{1-1}(\theta,\phi) \big] e^{- i \omega_\pm t} \, .
\end{equation}
Some algebra shows that the spatial/time dependence of these two modes are
\begin{align}
\label{eq:xi+}
    \delta F_+ &\propto \sin \theta \cos \phi \cos (\omega_+ t) \nonumber \\
    &\propto x \cos (\omega_+ t) \nonumber \\
    & \propto Y_{10,x} \cos (\omega_+ t)
\end{align}
and
\begin{align}
\label{eq:xi-}
    \delta F_- &\propto \sin \theta \sin \phi \sin (\omega_- t) \nonumber \\
    &\propto y \cos (\omega_- t) \nonumber \\
    & \propto Y_{10,y} \cos (\omega_- t)
\end{align}
Note also that the $m=0$ mode has
\begin{align}
\label{eq:xi0}
    \delta F_0 &\propto \cos \theta \cos (\omega_\alpha t) \nonumber \\
    &\propto z \cos (\omega_\alpha t) \nonumber \\
    & \propto Y_{10,z} \cos (\omega_\alpha t)
\end{align}
The extra subscripts on the $Y_{10}$ spherical harmonics $(x,y,z)$ refer to the axis with respect to which the spherical harmonic is defined.  Note that neither of the modes in Eqn.~\ref{eq:xi+} or Eqn.~\ref{eq:xi-} propagates around the equator like uncoupled $m=\pm 1$ modes. Instead, both modes are standing modes, aligned with the $x-$ and $y-$axes. The $m=0$ mode is a standing mode aligned with the $z-$axis. Thus, we find that strong tidal coupling naturally transforms an $\ell = 1$ dipole triplet of states to a set of $Y_{10}$ modes around three orthogonal axes.  Hence, tidally coupled $\ell=1$ modes can behave identically to the tri-axial pulsation behavior necessary to explain the observations of TIC 184743498.  

Tidal coupling will have a similar effect on $\ell=2$ modes, coupling the $Y_{21}$ and $Y_{2-1}$ modes into a perturbed doublet, and the $Y_{22}$, $Y_{20}$, and $Y_{2-2}$ modes into a triplet. This will produce more complicated spatial patterns than those discussed above, with different amplitude/phase modulation over the orbit. We plan to examine these signatures in future work and simultaneously search for these signatures in other stars in the {\it TESS} data.

As mentioned above, tidal coupling also couples modes of different $\ell$, which is not accounted for above. For weak tidal distortion, the $\ell=2$ component of the tidal distortion dominates, such that coupling between modes differing by $\Delta \ell = 2$ and $\Delta \ell=0$ is strongest. For stronger tidal distortion, the $\ell=3$ component of the tidal distortion becomes more important, enabling coupling between modes differing by $\Delta \ell = 3$ and $\Delta \ell=1$. In this case, the star and mode eigenfunctions will be asymmetric across the $x-y$-plane, such that pulsations can be strongly trapped on either side of the star (i.e., the side facing toward or away from the companion). This leads to the observed ``tidal trapping" or ``single-sided pulsator" phenomenon (e.g., \citealt{2020NatAs.tmp...45H,2020MNRAS.494.5118K}).

In general, tidal coupling could lead to complex behavior of tidally perturbed pulsations, and the effects of strong tidal distortion, centrifugal, and Coriolis forces should all be taken into account. However, the simple case shown above demonstrates a straightforward mechanism through which tidal coupling could induce pulsations to naturally align with the $x$, $y$, and $z-$axes of the star. More thorough calculations incorporating the effects listed above, and using larger networks of coupled modes, will be needed for a full understanding of pulsations of tidally distorted stars. 

\section{Summary, Conclusions, and Future Directions}
\label{sec:sumconclude}

In this work, we report the discovery and analysis of a tight eclipsing binary with at least nine tidally tilted pulsation modes, TIC 184743498.   

Evidence from our own ETV curve, as well as from Gaia's discovery of an astrometric acceleration solution, both indicate that there is a third star in the system in a wide orbit of thousands of days.  We have analysed the available archival data for this system to investigate the stellar parameters and the evolutionary state of the system using an SED fitting code. This code incorporates measured RVs from Gaia, stellar evolution models, and other parameters extracted from the light curve with the modeling code {\sc Lightcurvefactory}. We find that the pulsating primary star has $M_1 \simeq 1.83\,$M$_\odot$, $R_1 \simeq 1.72$\,R$_\odot$, and $T_{\rm eff,1} \simeq 8500\,$K, while the secondary has $M_2 \simeq 1.37$\,M$_\odot$, $R_2 \simeq 1.35$\,R$_\odot$, and $T_{\rm eff,2} \simeq 6870$\,K.  The primary star has evolved somewhat away from the ZAMS at an age of 460 Myr, while the secondary star is still on the ZAMS.  The uncertainties on the properties of the inferred third star are large, but it is likely not too different from the secondary star in the inner binary, and apparently contributes $\sim$10\% of the system light.  

We then analyzed the pulsations of TIC 184743498 in detail. The system exhibits nine pulsation modes, 5 of which have pulsation amplitude maxima at the binary eclipses and phase shifts of $\pi$ at the ELV peaks. We conclude that the first 5 pulsations can be explained by dipole modes $Y_{10}$ about a pulsation axis which has been tilted into the orbital plane and which lies along the tidal axis.  We call these `$Y_{10,x}$' modes.

Four other pulsations exhibit amplitude maxima at the ELV peaks and $\pi$ phase shifts at the eclipses. We have shown that while the latter modes may appear to be $Y_{11}$ modes based on their amplitudes and phases, this explanation is untenable because at the relatively low orbital inclination angle of this system ($65^\circ$), such doublets would have an unmistakable central peak in their echelle diagram. Such a central peak is a visual description that the mode amplitude is nonzero at $t_0$ (i.e., at the eclipses) for these modes; however, the observed mode amplitudes do, in fact, vanish at the eclipses. The lack of such a central peak shows that these modes are actually $Y_{10}$ about an axis we define as $y$, which is orthogonal to the tidal axis and the angular momentum axis. We term these `$Y_{10,y}$' modes.

We further found that one or both of two remaining singlet pulsation modes are also $Y_{10}$ modes, but about the angular momentum axis $z$, and we investigated this hypothesis with single-star pulsation models. We found that the expected pulsation spectrum of radial modes with a star of the parameters derived from the SED fit is significantly less dense than the observed pulsation frequencies. Thus, the singlet pulsation modes cannot both be radial pulsations, but rather at least one of them must be a $Y_{10,z}$ mode, i.e., about the z-axis. 

To study this hypothesis of three different pulsation axes, we explored a natural explanation in which strong tidal coupling can convert an otherwise uncoupled triplet of $\ell = 1$ dipole states to all $Y_{10}$ modes, but about three orthogonal axes. These axes are (i) the tidal axis, (ii) the orbital angular momentum (i.e., stellar rotation) axis, and (iii) the direction in the orbital plane perpendicular to the tidal axis.  This is exactly what we are seeing in the total set of 9 dipole modes and two singlet modes observed in TIC 184743498.  Additionally, we show that the transformed original $Y_{1\pm1}$ modes, which have circulating pulsation modulations, are reduced to standing waves.

Every newly discovered tidally tilted pulsator presents distinctly unique behavior in its pulsations, and challenges previously understood ideas about tidally tilted pulsations. TIC 184743498 contributes to this diversity as the first identified tri-axial pulsator.

More generally, the discovery reported here of a tri-axial set of pulsation axes has a wider application to the broad subject of stellar pulsations.
We have seen that simple modes (such as doublet-dipoles) are easier to understand in the context of tidal tilting than complex multi-element modes; the latter of which are, in contrast, far easier to spot in echelle diagrams.  Finding the numerous doublets in TIC 184743498, and their interpretation, have allowed us to identify and interpret the modes of nearly all the prominent pulsations in this star.

 With this in mind, we hope to search for more TTP candidate stars with simpler modes such as doublets rather than those with complex multi-element modes, such as HD\,265435 \citep{jayaraman22}.  Future observations of TTPs in different photometric bands may reveal new insights into mode cavities/propagation at different stellar layers, which, in turn, can provide an exquisite new view into stellar interiors. We also highlight the fact that toy models such as the one presented in this work should allow us to develop a preliminary understanding of previously-unexplainable phenomena, such as complicated mode couplings in stellar interiors. 

In this work we have also shown that, in systems with tidally tilted modes, part of the pulsation geometry can be constrained, in some cases rather tightly, by the determination of the orbital and system parameters, especially the orbital inclination angle. It was just this tilt of the orbit that allowed us to realize that we were dealing with tri-axial pulsations in TIC 184743498, otherwise, the pulsational multiplet structures would be different.  This orbital constraint can, and should, be applied to other TTP systems.  

This work also motivates us to reexamine other previously identified TTP candidates to see if any of their multiplets can be fit into this paradigm of tidally induced doublets.  In this regard, the simple model for tidal tilting presented here needs expansion to quadrupole modes.  Once the signatures of tidally tilted quadrupole modes, under the paradigm developed in this work, are understood, we will be able to search for these specific features in other pulsating stars in binaries. Given a small set of stars with precisely measured stellar parameters, and nearly complete sets of identified pulsation modes, detailed stellar modeling of these modes will be warranted. 

Finally, we note that the 200-second cadence data that will continue to come from {\it TESS}'s all-sky survey will be invaluable in helping us determine the relationship between the stellar pulsation axes and the strength of the tidal perturbation.  These data will also allow us to find all manner of new types of pulsations, and perhaps answer some of the lingering questions in the field.

\section*{acknowledgements}

%We are grateful to an anonymous referee whose comments and suggestions helped clarify the presentations in this paper.

This paper includes data collected by the {\it TESS} mission.  Funding for the {\it TESS} mission is provided by the NASA Science Mission directorate. Resources supporting this work were provided by the NASA High-End Computing (HEC) Program through the NASA Advanced Supercomputing (NAS) Division at Ames Research Center to produce the SPOC data products. Some of the data presented in this paper were obtained from the Mikulski Archive for Space Telescopes (MAST). STScI is operated by the Association of Universities for Research in Astronomy, Inc., under NASA contract NAS5-26555. Support for MAST for non-HST data is provided by the NASA Office of Space Science via grant NNX09AF08G and by other grants and contracts.

G.\,H.\,acknowledges financial support from the Polish National Science Center (NCN), grant no. 2021/43/B/ST9/02972.   This research was supported by the Erasmus+  programme of the European Union under grant number 2017-1-CZ01-KA203-035562.  

T.\,B.\, has received funding for this project from the HUN-REN Hungarian Research Network.

%{\bf Author contributions}  V.\,Z.~discovered the source and its tidally tilted pulsations, did the initial data analysis, produced the first echelle diagram and identified the $\sim$10 doublets, and is the main curator of the manuscript.  S.\,R.~produced the first reconstructed pulsations for this source, verified and refined the analysis of the pulsations, developed and used the SED fitting code to extract the system parameters, and worked extensively on the text.  R.\,J.~ developed the pipeline for the Summary Sheets of the TESS SPOC sources that were searched to find this object, ran statistical analyses of the MCMC posteriors, worked to develop the text.  D.\,K.~was the irst to infer that some of the pulsation modes had a pulsation axis in the y direction; this was key to our breakthrough in understanding the tri-axial nature of the pulsations. G.\,H.~constructed independent echelle diagram as well as amplitude and phase diagrams for this source, and produced the stellar pulsation models that confirmed one of the singlets must be a non-radial mode along the orbital axis; this finalized the conclusion that this object is indeed a tri-axial pulsator.  J.\,F.~came up with, and formalized, the mechanism by which the tidal forces couple two modes with a pulsation axis in the orbital angular momentum direction (z), into modes that pulsate along the x and y axes. T.\,B.\, ran his Lightcurvefactory code to analyze the TESS orbital lightcurve, and extracted the key parameters of the binary system.

\vspace{0.5cm}
\noindent
{\em Data availability}
\vspace{0.2cm}

\noindent
The {\it TESS} data used in this paper are available on MAST.  All other data used are reported in tables within the paper. The {\tt MESA} binary evolution `inlists' are available on the {\it MESA} Marketplace: \url{http://cococubed.asu.edu/mesa_market/inlists.html}.

\bibliography{184743498.bib}

%\end{thebibliography}
%===========================================================================================
%                                       APPENDICES
%==================================================================================

\appendix

\end{document}